\documentstyle[11pt,paspconf,epsf]{article}
\begin{document}

\renewcommand{\floatpagefraction}{0.7}
\renewcommand{\topfraction}{0.90}
\renewcommand{\bottomfraction}{0.90}

\def\kms{${\rm km\ s^{-1}}$}

\title{The Sunyaev-Zeldovich Effect as Microwave Foreground and Probe of
Cosmology}
\author{Gilbert P. Holder and John E. Carlstrom }
\affil{Department of Astronomy and Astrophysics, University of Chicago,
Chicago, IL 60637}

\begin{abstract}
The Sunyaev-Zel'dovich (SZ) effect from clusters of galaxies should
yield a significant signal in cosmic microwave background 
(CMB) experiments at small
angular scales ($\ell \ga 1000$). 
Experiments with sufficient frequency coverage should be 
able to remove much of this signal in order to recover the primary
anisotropy. The SZ signal is interesting in its own right;
the amplitude and angular
dependence are sensitive to both cosmology and the evolution of the
gas. Combining CMB measurements with planned non-targeted
SZ surveys could isolate the cosmological effects, 
providing CMB experiments with a low-redshift test of cosmology
as a consistency check. Improvements in the determination of the
angular diameter distance as a function of redshift from 
SZ and X-ray observations of a large sample of clusters will 
also provide a probe of cosmology.

\end{abstract}


\section{Introduction}
\label{sec:introduction}

The Sunyaev-Zel'dovich (SZ) effect has been predicted for several
decades (Sunyaev and Zel'dovich 1972), but it has only recently become
a valuable observational tool. In the past few years, we have seen
the observational programs go from ones aimed at detecting the effect
to using it to map out massive clusters of galaxies
(e.g., Birkinshaw 1999). In many ways, this
is similar to the rapid evolution of CMB anisotropy experiments.

The SZ effect arises from hot gas between us and the surface of last
scattering distorting the thermal spectrum of the CMB. This could
describe a large variety of situations, but the SZ effect is usually
discussed in the context of the effects of low redshift (below $z\sim  10$)
gas that is heated mainly by gravitational infall in the course of structure
formation. That will be the focus of this paper as well.

There are two SZ effects which are discussed. The first is the so-called
thermal effect arising from Compton scattering, while the second is
known as the kinetic effect, arising from simple 
Thomson scattering by a moving object.

The cool CMB photons traversing hot gas can undergo Compton scattering
from the hot electrons, gaining a small amount of energy in the exchange.
A typical fractional energy gain by the photons, 
in the non-relativistic limit of low electron and photon 
energies would be roughly ${kT_{gas} \over m_e c^2}$ per collision. 
The cross-section for collision is simply the Thomson cross-section,
$\sigma_T$, leading to an optical depth for a gas of electron number
density $n_e$ and characteristic size $L$ of $\tau \sim n_e \sigma_T L$.
Thus, the CMB photons as a whole should get a fractional energy gain
roughly given by the Compton $y$ parameter, defined as the fractional
energy gain multiplied by the optical depth to scattering. 

\begin{figure}
\epsfysize=3.7 in
\centerline{\epsfbox{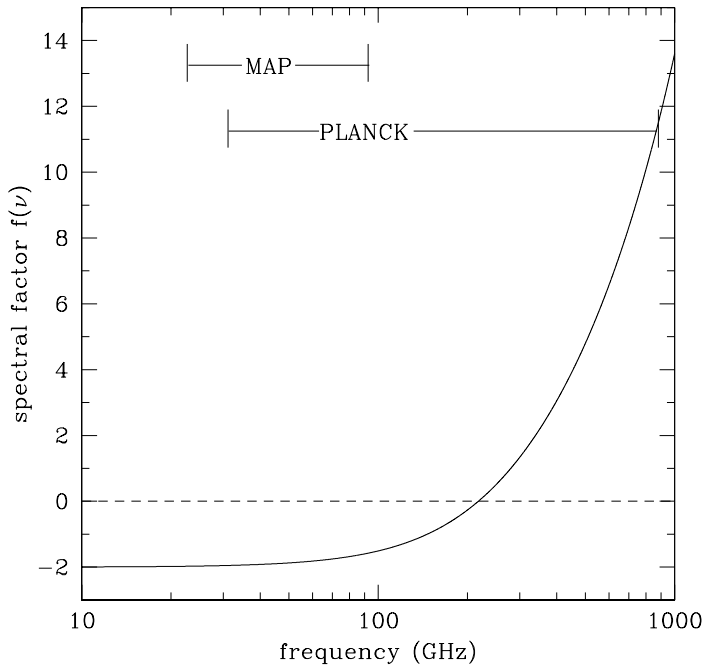}}
\caption{Spectral behavior of thermal SZ effect, where ${\Delta T \over
T_{CMB}}= f(\nu)y$,
as a function of frequency, in the non-relativistic limit.
Note that the high frequency area is far out in the Wien part
of the CMB blackbody curve, and therefore there is very little
signal to use to detect a temperature difference.}
\label{spectrum}
\end{figure}

This scattering conserves photon number, so the net effect is to shift
the CMB spectrum up in energy, distorting the blackbody curve slightly.
This leads to a decrement at low frequencies, as more photons have been
scattered out of those energy bins than have been scattered in,
while there is an increment at high frequencies. 
The exact spectral form thus depends on the slope of the blackbody
curve. The crossover
frequency is near $217$ GHz, with small relativistic corrections
(Rephaeli 1995) making the crossover frequency a weak function
of the gas temperature.
This is a distinctly non-thermal spectrum, and CMB experiments with
good frequency coverage will be able to easily distinguish this
from purely thermal fluctuations.
The spectral behavior is shown in Figure \ref{spectrum}. In particular,
note that the spectral behavior is slowly varying over the entire MAP
CMB satellite experiment
frequency coverage (22-90 GHz), and thus could be a difficult
foreground to remove. On the other hand, the PLANCK satellite will
have coverage from 30-850 GHz, allowing good sampling of the peculiar
spectral behavior of the thermal SZ effect.

The kinetic effect arises from a moving mass of hot gas between us and
the (mis-named in this case) surface of last scattering. The moving object
sees a significant dipole in the CMB, with a hot spot in the direction
of its motion. Thomson scattering will remove this dipole from the
photons that undergo scattering. Back in the rest frame of the CMB, the
gas will have created an apparent hot spot for observers that see the 
gas moving toward them and a cool spot for observers with the gas moving
away from them.

The kinetic effect does not cause
a deviation from a blackbody spectrum; it rather leads to a blackbody
with a slightly different temperature. This means that the kinetic SZ
effect has exactly the same spectral behavior as the primary CMB anisotropy.
Thus, the kinetic effect is a foreground which is impossible to remove
by spectral methods and it is important to understand its expected
contributions to CMB anisotropy. Likewise, the primary CMB anisotropy
can be a major source of confusion for determining peculiar velocities
of clusters using the kinetic SZ effect.

The kinetic effect involves an energy shift per scattering which is 
effectively just a redshift, and thus on the order
of $(v_{bulk}/c)$. Bulk peculiar velocities of clusters are expected
to be less than $\sim 1000$ \kms in any cosmology. 
The thermal effect depends on the temperature of the electron gas,
with an energy shift per scattering on the order of ${kT_e \over m_e
c^2}$.
Assuming that the electrons
and ions share the same temperature (Fox and Loeb 1997),
the thermal energy in the gas is typically on the order 
$m_H v_\sigma^2$, where $v_\sigma$ is the internal velocity dispersion
in the potential well of the cluster, which can be $\sim1000$ \kms.
At most frequencies, the
thermal effect will be larger than the kinetic by a factor of ten or
more.
 
The peculiar spectral behavior
of the thermal effect means that the relative effects of the kinetic
and thermal effects change with frequency, with the only SZ signal
at the null of the thermal effect coming from the kinetic effect. It
has been suggested that the peculiar velocity field
of clusters could be mapped out in this way, 
but the small magnitude of the kinetic SZ effect makes this
a very difficult experiment. Upper limits have been
placed on the peculiar velocities of clusters (Holzapfel {\em et~al.}\ 1997a),
but a clear detection of the kinetic effect has not yet been
observed.

\section{Clusters of Galaxies}
\label{sec:clusters}

Clusters of galaxies are excellent sources of the SZ
effect. Most of the baryonic mass of these objects is in the form
of diffuse inter-galactic gas, heated to the virial temperature of
the cluster. With cluster masses as large as $10^{15} M_\odot$
and radii of several Mpc, the gas can be $\sim 10$ keV.

It is a fairly good approximation to take clusters as isothermal spheres
with a truncation at the virial radius:
\begin{equation}
\rho(r) = {\rho_\circ \over 1+({r \over r_c})^2  } \quad {\rm for} \quad r<R_v 
\end{equation}
where the virial radius corresponds to the radius at which the infalling
gas is shocked and converts much of its kinetic energy into thermal
energy.
 
Assuming that the electron number density is also of this form, 
and that isothermality is a good approximation
(Markevitch {\em et~al.}\ 1998), then
we can derive a value for the central SZ effect:
\begin{equation}
{\Delta T_{\circ} \over T_{CMB}} = 2 \, f(\nu)\, {kT_{gas} \over m_e c^2}\, \sigma_T 
 \, n_\circ  \, r_c  \, \arctan{({R_v \over r_c})}
\end{equation} 
where $f(\nu)$ in the above gives the spectral dependence, approaching -2 in
the Rayleigh-Jeans limit.

The SZ central decrement (at low frequencies) depends on the distribution
of electrons in the cluster, with more condensed clusters having a stronger
signal for a given mass. 
For a massive galaxy cluster, $n_\circ \sim 0.01 \, {\rm cm^{-3}},
T \sim 10 \, {\rm keV}, r_c \sim 250 \, {\rm kpc}$, and the virial radius is
usually a factor of ten (or more) 
larger than the core radius. Using these numbers,
we arrive at a central decrement of $\Delta T/T_{CMB} \sim 10^{-3}$. 
However, the central density and core radius are expected to vary from
cluster to cluster in a poorly understood way, making the central decrement
a difficult observable to understand.

The central decrement is a redshift-independent probe of clusters. The
magnitude of the central decrement is sensitive to the degree of 
concentration of the cluster, as well as the path length along the
line of sight through the cluster. For a given set of cluster properties,
though, the central decrement does not depend on the redshift.

An observable that is much easier to understand is the total integrated
SZ signal:
\begin{equation}
S \equiv S_{\circ} \int d\Omega \,{\Delta T \over T_{CMB}} = S_{\circ} \int d\Omega \int dl \, f(\nu)
{kT_{gas} \over m_e c^2} \sigma_T n_e(l)
\end{equation}
where $S_{\circ}$ gives the conversion to physical units such as Jy. 
Writing the integral over solid angle as $dA=d\Omega D_A(z)^2$, with $D_A(z)$
the angular diameter distance, this becomes
\begin{equation}
S = S_{\circ} f(\nu) {kT_{gas} \over m_e c^2} \sigma_T {N_e \over D_A(z)^2}
\end{equation}
with $N_e$ the total number of electrons, which should be simply related
to the total mass. Note that we have assumed nothing about the distribution
of the mass, only that the gas is isothermal. Any experiment with a 
beam size larger than the cluster will not be sensitive to the degree of
concentration of the gas. This is not the case for most observables of
a system. For example, X-ray bremsstrahlung emission is proportional
to $n_e^2 T^{0.5}$, making the integrated emission highly sensitive to the
degree of concentration of the gas.

Finally, it is important to note that the SZ effect, as a distortion
of the blackbody spectrum, is independent of redshift for a given
set of cluster properties and therefore all redshifts contribute equally to the
magnitude of the effect along a given line of sight. Of course, a 
nearby cluster will intersect more lines of sight, and thus will have
a larger integrated effect over the sky, as seen by the inclusion of
the angular diameter distance in the expression for the total integrated
effect. 
Clusters are typically a few arcminutes in extent, so
as long as the beam size is $\la 1'$
the SZ effect is essentially independent of redshift. Even in the case
of large beams, the redshift dependence is not particularly restrictive.
For comparison, the angular diameter distance is 
smaller than the luminosity distance by a factor of $(1+z)^2$.

\section{Observations}
\label{sec:observations}

As for observations of intrinsic anisotropy in the CMB, observations
of the SZ effect have progressed over the last several years from low
S/N detections and upper limits to high confidence detections and
detailed images. The dramatic increase in the quality of the
observations is due to improvements both in low-noise detection
systems and in observing techniques, usually using specialized
instrumentation with which the systematics that often prevent one
from obtaining the required sensitivity are carefully controlled. Such
systematics include, for example, the spatial and temporal variations
in the emission from the atmosphere and the surrounding ground.

The SZ effect observations so far have been targeted mainly
toward X-ray luminous galaxy clusters. Large scale
non-targeted surveys for the SZ effect could, in principle,
provide a unique and powerful probe of the distant universe
as discussed in section~\ref{sec:non-targeted} The power of such
surveys is
due to the independence of the SZ effect on the redshift
to the galaxy cluster or filament causing the scattering. 
Unfortunately, the 
sensitivity of the current instrumentation used to measure the effect is not
high enough to allow a large region to surveyed in a reasonable
amount of time. This situation may change soon, as instruments
with the required sensitivity are being planned now. 

The existing observations of the SZ
effect toward galaxy clusters can be roughly divided into
two classes:
1) single dish observations toward low 
redshift galaxy clusters  ($z < 0.2$) at resolutions of several 
arcminutes, and 2) interferometric observations of distant
clusters ($z > 0.2$) at resolutions of roughly an arcminute and higher.
A recent review of the observations can be found in 
Birkinshaw (1999). Here we discuss briefly a few of the
results to provide the reader with a measure of the
quality of the data presently available.

Motivation for obtaining measurements of the SZ effect toward 
galaxy clusters is provided by the ability to combine
SZ effect and X-ray emission data to determine cluster distances, 
and therefore the
Hubble constant, and also to determine the fraction of
the total cluster mass contained in the hot gas.  
The distance measurements are compelling as the estimates
do not share any of the systematics inherent in more traditional
methods; the technique relies solely on an understanding
of the cluster gas. The SZ effect distances do, of course,
suffer from other systematics. The gas mass fraction determinations
are compelling as
the gas mass accounts for most of the known baryonic mass
of the cluster (Forman and Jones 1982, White {\em et~al.}\ 1993) 
and therefore the gas mass fraction allows a
good measure of the baryonic mass fraction for the cluster.
Furthermore, the baryonic gas mass fraction for massive clusters should
reflect the universal value $\Omega_B / \Omega_M$. Thus,
we can solve for $\Omega_M$ by using the cluster gas mass
fraction and the value of $\Omega_B$ derived from
big bang nucleosynthesis (BBN)
calculations (Copi, Schramm, and Turner 1995)
constrained to fit the observed 
primordial abundance of light elements (e.g., Burles and Tytler 1998). 

As discussed in Birkinshaw (1999), the first measurements of
the SZ effect were made with single dish radio telescopes. 
Successful detections were obtained, although the reported
results show considerable scatter, reflecting the difficulty of the
measurement (see earlier review by Birkinshaw 1991 and references therein).
These observations illustrated the need for telescopes and receiver
systems designed explicitly to minimize differential atmospheric
emission and ground pickup. Recent state-of-the-art single dish
observations at radio (Herbig {\em et~al.}\ 1995, Myers {\em et~al.}\ 1997) and at millimeter wavelengths
(Wilbanks {\em et~al.}\ 1994, Holzapfel {\em et~al.}\ 1997b) have resulted in significant detections of the effect
and limited mapping. Observations near the null of the thermal
effect have been made in an attempt to measure the peculiar velocities
of galaxy clusters. An upper limit of about 2000~\kms\ has been 
set for the $z\sim0.2$ clusters Abell~2163 and Abell~1689
by Holzapfel {\em et~al.}\ (1997a), who also report a significant detection
of the increment at frequencies above the null (see also
Lamarre {\em et~al.}\ 1998).

\begin{figure}
\plotone{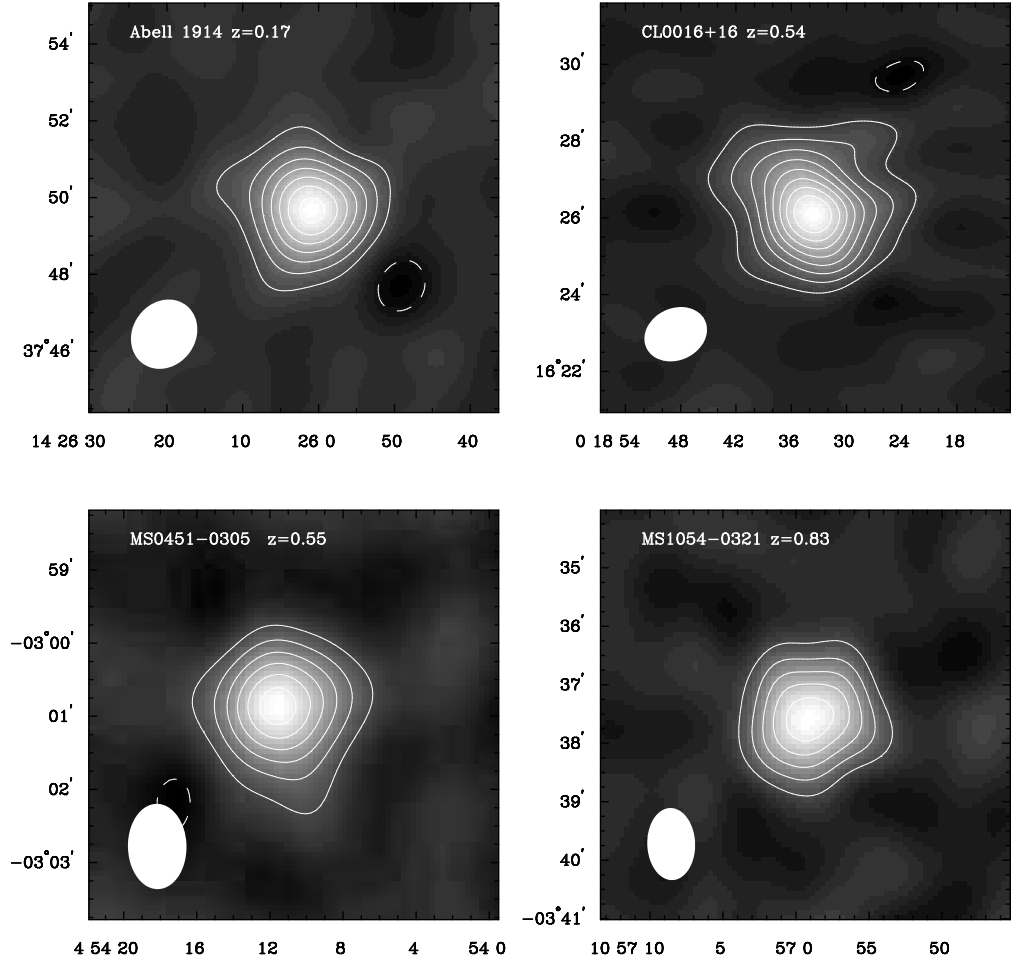}
\caption{Images of the the SZ effect observed toward four
galaxy clusters with similar luminosities, but spanning redshifts from 0.17 to 0.83. For
all four clusters the 
contour levels are integer multiples of --75~$\mu K$,
which corresponds to
1.5 to 5 times the noise level, depending on the cluster.
The solid white ellipse in the bottom left corner of 
each panel represents the effective resolution of the image. 
The comparable SZ decrement observed for each cluster 
illustrates the independence on the SZ effect signal on redshift.
The data were taken with low-noise cm-wave receivers installed 
on the OVRO and BIMA mm-wave interferometric arrays. }
\label{fig:szimages}
\end{figure}

Interferometric techniques have been used to produce high
quality images of the SZ effect (e.g., Jones {\em et~al.}\ 1993,
Grainge {\em et~al.}\ 1993, Carlstrom, Joy, and Grego 1996,
Grainge {\em et~al.}\ 1996,
Carlstrom {\em et~al.}\ 1998). The high stability of
interferometry is being exploited to
make these observations. Interferometry
also provides data on the cluster over a large
range of angular scales making it possible to 
remove contaminating emission from individual radio sources.
Lastly, it is possible to produce two-dimensional
images of the SZ effect with interferometry. 
We show 
in Figure~\ref{fig:szimages} four of the 27 clusters
imaged with the cm-wave SZ effect experiment on
the OVRO\footnote{An array of 10.4 m mm-wave telescopes
located in the Owens Valley, CA and operated by Caltech.}
and BIMA\footnote{An array of
ten 6.1 m mm-wave telescopes located
at Hat Creek, California and operated by the 
Berkeley-Illinois-Maryland-Association} interferometers 
(Carlstrom {em et~al.}\ private
comm.).

In his recent review of the SZ effect,
Birkinshaw (1999) provides a summary of 
Hubble constant determinations from the above single
dish and interferometric, as well as other, SZ effect 
observations of clusters with 
redshifts as high
as 0.55. The mean value of the derived Hubble constants 
is $\sim 60 \ {\rm km\,s^{-1} Mpc^{-1}}$ with no
meaningful constraint on the deceleration parameter. 
The scatter in the Hubble constant
measurements is of order $20\ km s^{-1} Mpc^{-1}$. 
Birkinshaw
points out, however, that 
many of the observational uncertainties, such
as the X-ray and radio absolute flux scales are correlated between
the observations.

Two groups have reported cluster gas fractions 
based on SZ effect data
(Myers {\em et~al.}\ 1997, Grego 1999, Grego {\em et~al.}\ 1999).
Myers {\it et al.}\ use their single dish SZ effect observations
combined with $\beta$-models for the gas density distribution 
constrained by X-ray imaging data, and with gas temperatures
determined by X-ray spectroscopy, to determine the
gas mass fractions $f_g$ for a sample of four clusters.
They report $f_g h = 0.061\pm 0.01$
for three nearby clusters, although a fourth
cluster in their sample gives $f_g h = 0.166 \pm 0.014$.
Grego {\it et al.}\ report the results for
18 clusters imaged with the OVRO and BIMA cm-wave 
interferometric SZ system. They use the SZ effect data itself
to constrain the $\beta$-models and X-ray spectroscopy
for the gas temperatures. For the entire
sample spanning redshifts from 0.171 to 0.826
and assuming an open $\Omega = 0.3$ cosmology, they 
find $f_g h = 0.077 ^{+0.006}_{-0.010}$
scaled to $r_{500}$, the radius at which the
overdensity of the cluster is 500. Using only clusters
with redshifts less than 0.25 so that the
cosmology assumed will have little effect on
the gas mass fractions derived, they find $f_g h = 0.085 ^{+0.011}_{-0.145}$,
again scaled to $r_{500}$. 

Assuming that the baryonic mass fraction for clusters
equals the universal value and using
$\Omega_B h^2 = 0.019 \pm 0.001$ determined from
BBN and primordial abundance measurements (Burles and Tytler 1998),
these gas mass fractions imply $\Omega_M h \sim 0.25$.

Future improvements in SZ effect observations, such
as dedicated interferometric arrays and single dish
telescopes equipped with bolometric arrays, will
allow detailed imaging of many more than the
order 30 clusters imaged to date and should
allow definitive measurements of the Hubble constant
and deceleration parameter. Commensurate improvements
in the absolute calibration scale for both the
radio and X-ray observations are also needed
and will be provided, in part, by the AXAF/Chandra satellite 
at X-ray wavelengths, and 
by careful ground based efforts and by the 
planned CMB satellite missions at radio wavelengths.
The expected increase
in sensitivity and imaging speed will make large
scale non-targeted surveys of the SZ effect possible,
allowing a complete inventory of clusters at high
redshifts.

\section{The SZ Effect at Higher Redshifts}
\label{sec:high_redshift}

The imprint of distant clusters on the CMB has two important implications.
As the SZ effect is the upscattering of CMB photons, it should come as no
surprise that this can be a contaminant in CMB experiments. The thermal SZ
effect can become larger than the CMB signal around $\ell \ga
2000$, depending on the cosmology and cluster properties.  
The thermal effect
has a distinctive frequency signature, allowing a multi-frequency experiment
with sampling on both sides of the null (near 217 GHz) to separate it from
intrinsic CMB anisotropy. 
The kinetic SZ effect cannot be spectrally distinguished from the primary CMB
anisotropy, and could be impossible to remove from the data.
Fortunately, the kinetic signal is expected to be 
much lower than the thermal signal, and
thus never becomes dominant. 

While Planck has frequency channels that
trace out the thermal SZ spectrum on both sides of the null, this
is not the case for MAP which only has frequency coverage up to 
90 GHz. For the thermal SZ effect, this is still fairly far into the
Rayleigh-Jeans region, where the signal becomes independent of 
frequency. Thus, there is very little leverage for inferring the
amount of SZ contamination in measurements of the CMB. Several
papers have investigated the expected anisotropy due to 
high-redshift clusters (Cole and Kaiser 1989,
Atrio-Barandela and Muecket 1999,
Colafrancesco {\em et~al.}\ 1994). We discuss the contribution of
the SZ effect to the angular power spectrum in 
section \ref{sec:angular}

The second major implication is that surveys could be done to search
for this effect to map out the cluster abundance as a function of
redshift. This is expected to be a powerful probe of cosmology
(e.g., Barbosa {\em et~al.}\ 1996, Colafrancesco {\em et~al.}\ 1997) 
as discussed in section~\ref{sec:non-targeted} 

However, both the level of CMB contamination and the evolution of the
cluster abundance depend on cosmology and cluster physics. It is expected
that the dependencies are different for the two observables, and combining
CMB measurements with deep surveys should together separate cosmological
evolution of the number density from evolution of the gas properties.

In this section, we will lay out the tools required to probe
high-redshift clusters. Both the comoving number density and
the physical properties of clusters are expected to evolve
with redshift.  The number density evolution of clusters of
a given mass is fixed by the cosmology, while a model is
needed to relate mass and temperature as a function of redshift.
Beyond this, X-ray studies indicate that the distribution of gas
within clusters also evolves with mass and redshift,
and must
also be taken into account. We will deal with these points,
in turn, below.

\subsection{Cosmological Evolution of Cluster Abundances}
\label{sec:cosmological}

The prospect of using the SZ effect to detect high-redshift clusters
is very exciting. Presumably, an SZ experiment should be able to
find clusters at any redshift, given high sensitivity and large sky
coverage. The expected number of clusters at a given sensitivity
depends both on the cosmology and the history of the cluster gas.

In a low-density universe, structure stops forming when the expansion
starts to be driven by either curvature or a cosmological constant.
If we live in a 
low-density universe, the structure that we now see, by and large,
also existed out to a redshift of $z \sim 1$ and should be detectable
out to those redshifts. 
In a critical-density universe, the structure that
we now see probably formed very recently, and thus would not be
detectable at high redshift. 
This means that a given patch of sky should
contain more clusters in a low-density universe. 
Since the
SZ effect is redshift-independent, all of these clusters should
be detectable.

To illustrate this, we will show results generated by assuming that
the Press-Schechter (Press \& Schechter 1976) formula for the
evolution of the number density with redshift is correct. We will
adopt two models with one having $\Omega_M=1.0,\Omega_{\Lambda}
=0.0, h=0.5, \sigma_8=0.6$ and the other having $\Omega_M=0.3,
\Omega_{\Lambda}=0.7,h=0.6,\sigma_8=1.0$. 
and the usual CDM shape parameter
for the power spectrum was taken to be $\Gamma=0.25$.
Both of these models reproduce the local cluster abundance 
(Viana and Liddle 1999), within the rather large errors.

All work was done assuming a global baryon fraction $f_b = 0.2$.
This is in reasonable agreement with observations of X-ray emissions 
(Evrard 1997, Mohr, Mathiesen, and Evrard 1999) and SZ decrements 
(Myers {\em et~al.}\ 1997, Grego {\em et~al.}\ 1999) from clusters.
Comparison between expected yields from BBN 
(Copi, Schramm, and Turner 1995) and abundances of light elements 
requires that $\Omega_b h^2=0.02$ (Burles and Tytler 1998). While 
the assumed baryon fraction is consistent with BBN for $\Omega_M=0.3$, 
this is
not the case for $\Omega_M=1$. Thus, our results may overpredict
the expected SZ signal in high-density universes, if BBN is correct and
the measured baryon fractions are in error.

The Press-Schechter formula gives the number density of virialized objects
of a given mass,
specifically excluding any mass which is not virialized. 
This would be missing SZ signal from structures which have been significantly
heated but are not yet virialized. Thus, the results here are most likely to be underestimates of the total SZ signal from clusters, although
the importance of this effect has yet to be convincingly demonstrated.

For all of the results shown here, a lower mass limit of $10^{12} M_{\odot}$
was chosen, since masses below this would probably have cooled on a timescale
much shorter than a Hubble time. The reason for this is twofold. First of
all, low-mass objects are cooler to start with, and thus have less 
energy to radiate. Secondly, cooling becomes increasingly efficient per
unit mass at lower temperatures. 

\subsection{Mass-Temperature Relation}
\label{sec:mass-temp}

Press-Schechter gives a prediction for the number densities of objects
of various masses, while the SZ effect is sensitive to both the number
of electrons along the line of sight and the gas temperature. 
A conversion between mass and SZ signal is therefore required. 

The gas temperature can be obtained
by assuming that the gas is in virial equilibrium. 
Virial equilibrium requires that $kT \propto {G M_v \over R_v}$. The
exact constant of proportionality depends on the distribution of mass
inside the cluster.

We need an estimate of the virial radius to get a mass-temperature
relation. This is usually done by assuming the spherical collapse
model. 
A region that is overdense relative to the
critical density will eventually pull out of Hubble flow, turn around
and collapse. The time of turnaround will be at exactly half the time
of eventual collapse. Assuming that the epoch of observation is the
epoch of formation, the turnaround radius of an observed cluster can
be easily calculated given a cosmology. Ignoring $\Lambda$, the
virial relation asserts that $2K=-W$,
where K and W are the total kinetic and gravitational potential
energy, respectively. Given that $K=0$ at turnaround
and energy is conserved, this indicates that $W_v=2 W_{ta}$
(where $ta$ subscripts indicate properties at turnaround). 
This directly leads to $R_v={R_{ta} \over 2}$. Including the effects
of $\Lambda$ modify this relation slightly (Lahav {\em et~al.}\ 1991). 

At the same time, the
background universe has been expanding, changing the relative overdensity
of the bound region.  The mean overdensity of the bound region thus
depends on the cosmology.  
Fitting functions are given by Bryan and Norman (1998) for the mean overdensity
relative to the critical density, with their fit for a universe with
$\Omega_M+\Omega_{\Lambda}=1$ given by
\begin{equation}
\Delta(z) = 18 \pi^2 + 82 x -39 x^2
\end{equation}
where $x=\Omega_M(z)-1$. 

The viral relation tells us that 
$kT_{gas} \propto GM/R_{vir}$, but the exact constant of proportionality
depends on the distribution of matter. For this, we turn to numerical
simulations. Using the normalization of Bryan and Norman (1998), and
a notation where $H(z) \equiv 100h \, E(z)$, we use
\begin{equation}
\label{eqn:mass-temp}
T_{gas} = 1.4 \times 10^7 (h^2 \Delta(z) E(z)^2)^{1 \over 3}
({M \over 10^{15} M_\odot})^{2 \over 3} \, \, {\rm K}
\end{equation}

For a critical density universe, equation \ref{eqn:mass-temp} becomes
very simple, with  the temperature for a given mass
scaling simply as $(1+z)$. This is easy to understand, since all
lengths (including the virial radius) will scale as $(1+z)^{-1}$,
so $GM/R_{v} \propto (1+z)$. 

\subsection{Evolution of Cluster Gas}
\label{sec:evolution}

Evolution of cluster gas must also be taken into account. The
sizes of clusters are expected to evolve with the redshift of
formation,
with higher temperatures, and
therefore a stronger SZ signal for a given mass at higher 
redshift. For the total SZ effect flux
for a cluster, this gets diluted
by the decreasing angular extent, but the central decrement suffers
no such dilution. In fact, since the central decrement 
$\propto \rho r T_{gas}$, and $T_{gas} r \propto M$, the central decrement
for a fixed mass would be expected to nearly scale with the average
gas density.

If clusters evolve self-similarly, the central decrement 
(for $\Omega_M=1$) evolves as
\begin{equation}
\Delta T_o \propto kT_{gas}n_o r_c \propto M(1+z)^3 \quad ({\rm self - similar})
\end{equation}
since $kT \propto GM/R$ and $n_o \propto (1+z)^3$.

The gas doesn't have to evolve this way, and X-ray observations seem
to indicate that it does not
(Kaiser 1991, Mohr, Mathiesen, and Evrard 1999). 
For example, if reionization dumped
a substantial amount of entropy into the gas, or if the gas gained
a significant amount of entropy during infall, it is reasonable that
the core may not be able to adiabatically contract to form a compact core, 
and the core could be more diffuse than self-similarity would predict.
The outer regions should be largely unchanged, giving a relation for
the core radius:
\begin{equation}
r_c^2 \propto {\rho_{crit}(z)\over \rho_c} R_v^2
\end{equation}
The self-similar solution has $\rho_c \propto \rho_{crit}$, and
$r_c \propto R_v$. The core properties in that case are independent 
of cluster mass. If there is instead a minimum core entropy, then a 
gas with polytropic index 5/3 will have $\rho_c \propto T^{3 \over 2}$. 
The core properties now depend on cluster mass, with lower masses having
less concentration.
Under these conditions, we obtain a different relation for the 
central decrement:
\begin{equation}
\Delta T_o \propto M \rho_{crit}(z)^{0.5} T_{gas}^{3/4} \propto
M^{3/2} (1+z)^{9/4} \quad ({\rm constant\,entropy\,core})
\end{equation}
for a fixed mass and a flat matter-dominated cosmology. 

This is significantly different, both in its
mass dependence and its redshift evolution. Thus, we would expect
both qualitative and  quantitative differences in counts above
a given central decrement in the two gas evolution scenarios.

\begin{figure}
\centerline{
\epsfxsize=.51\textwidth \epsfbox{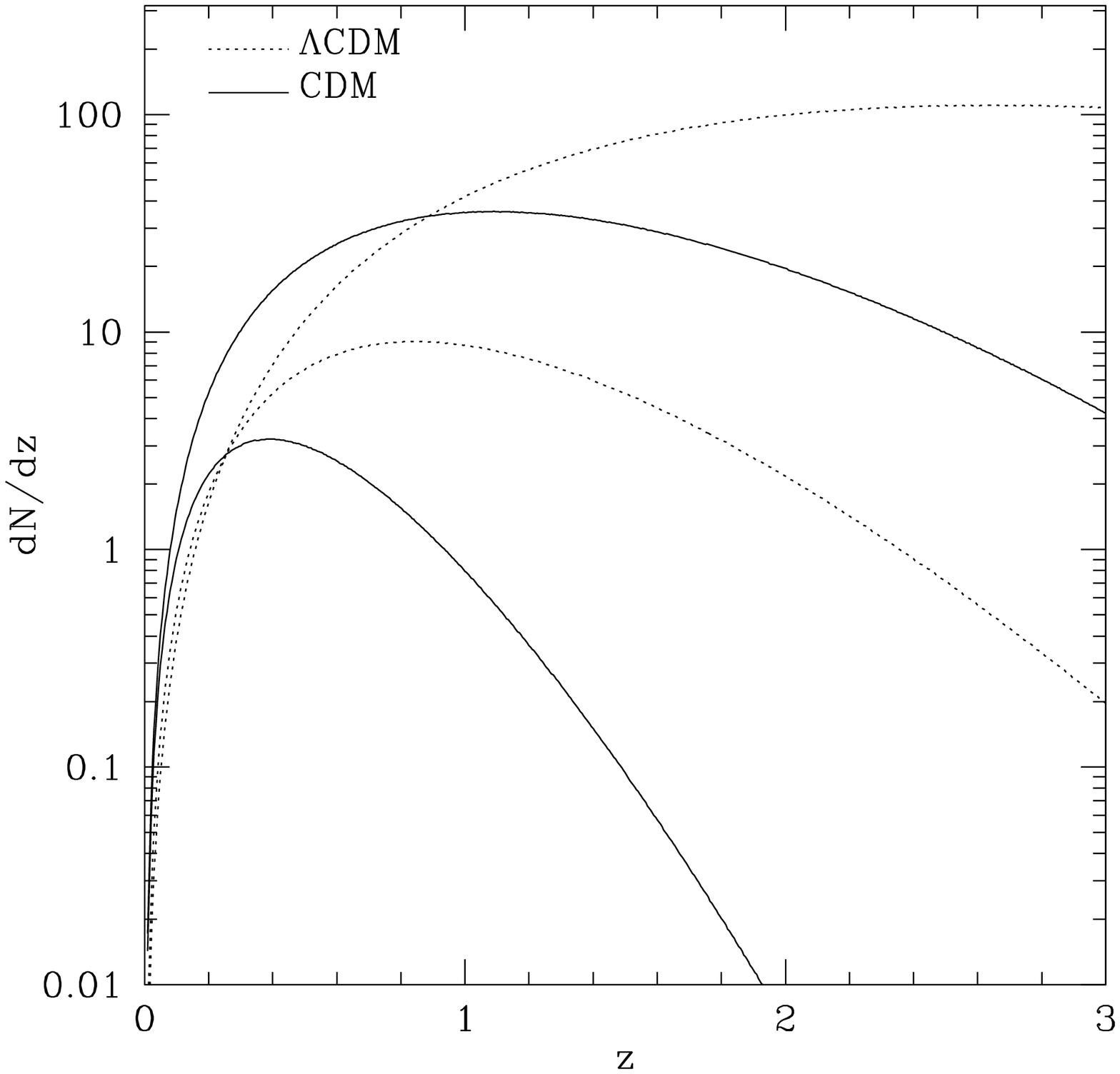} \hfil
\epsfxsize=.51\textwidth \epsfbox{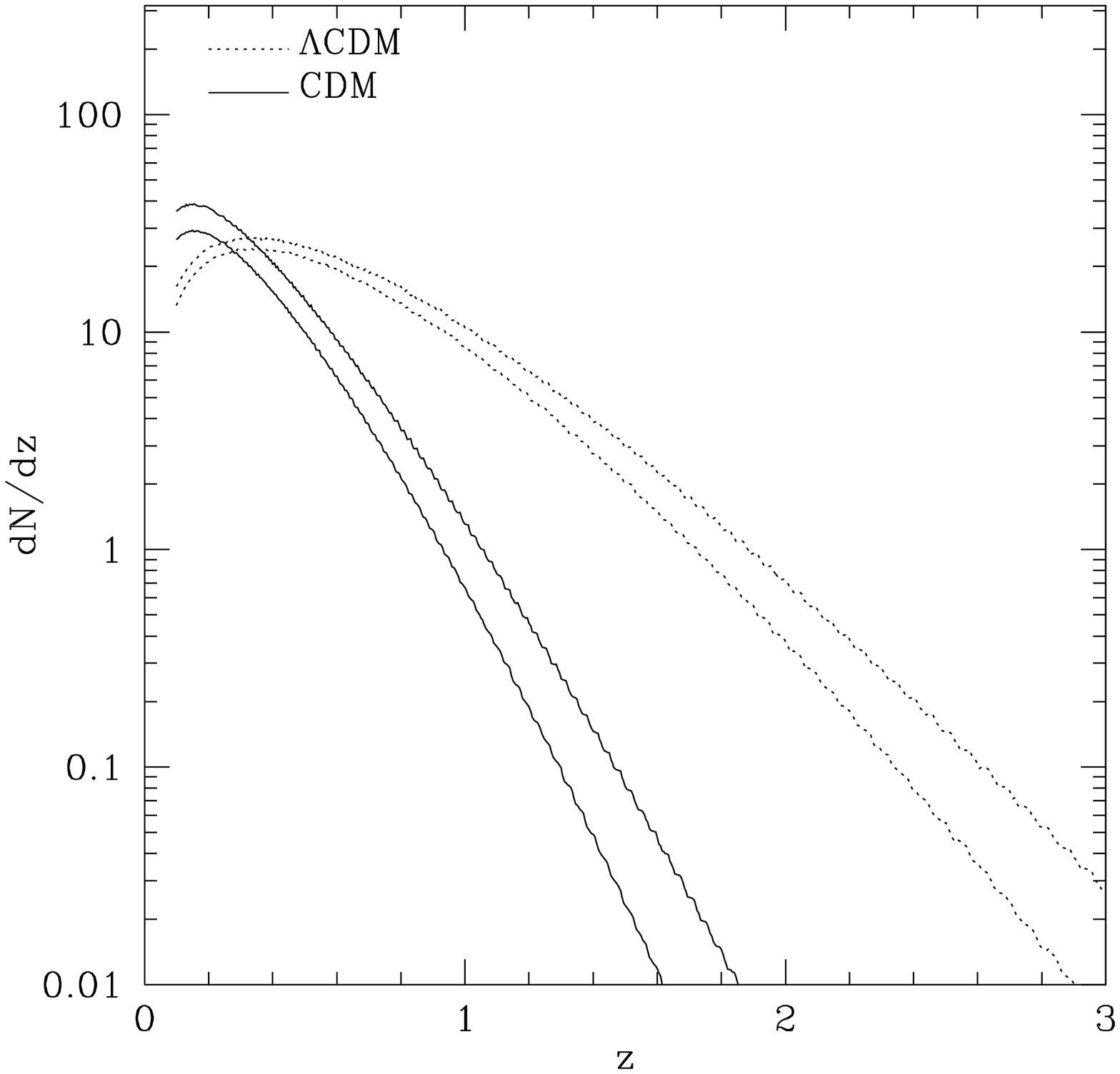}}
\caption{Counts per square degree of clusters with (left) central decrements
above a threshold of $100 \mu K$ (at 30 GHz)
and (right) total flux density above 0.5 mJy. The upper curve (at high
z) for each cosmology corresponds to self-similar evolution while 
the lower curve is for
evolution with a constant entropy core.}
\label{dndz_cd}
\end{figure}

This is shown in the left panel of Figure \ref{dndz_cd}. In this figure, 
we compare counts of clusters above a central decrement for self-similar 
scaling with a core:virial radii ratio of 1:10 and a constant entropy core
model normalized such that a 9 keV cluster has a central electron
number density of $2.0 \times 10^{-3}\,{\rm cm^{-3}}$. Counts are shown per
square degree for both a flat matter-dominated model and a model
with a cosmological constant. All calculations are done assuming
an observing frequency of 30 GHz.

The two models for evolution form an envelope for each cosmology, probably
bracketing the real evolution the gas follows. It can be seen that the
range is enormous, with gas evolution equally important to the cosmological
evolution. 

Compare this with the right panel of Figure \ref{dndz_cd}, where we show 
the counts per square degree for the same models, except now for all 
clusters with $S>0.5$\,mJy. Recall that it was earlier stated that the total
flux is independent of the gas distribution, assuming a constant
baryon fraction, and isothermality. 
The constant entropy core model does not have a constant 
baryon fraction, with low-mass clusters having 
a lower baryon fraction. This is easy to understand, as low-mass
clusters have shallower potential wells, making it more difficult
for pre-heated gas to fall in.
Even for this case, it can be seen that
the counts do not differ substantially from those expected from
self-similar evolution, and cosmology is clearly the defining 
element.

\section{Probing High-Redshift Clusters}
\label{sec:probing}

Using the tools of the previous section, we can now estimate the
expected signatures of high-redshift clusters. These can be
studied in at least two ways: non-targeted SZ surveys and
through the angular power spectrum of the anisotropy imprinted
on the CMB. The next few years will see both of these methods
implemented.

\subsection{Non-Targeted SZ Surveys}
\label{sec:non-targeted}

Current observations of the SZ effect have been targeted at 
clusters which were known to have significant X-ray emission.
Non-targeted surveys hold much promise for studying the evolution
of the cluster abundance.
Planned non-targeted SZ surveys are sensitive to both cosmology and gas
evolution. 
Optical follow-up of SZ surveys will be conducted to obtain the redshift
distribution.
This redshift
distribution is vital for finding the epoch of structure
freeze-out, which is set by the redshift at which either curvature or
$\Lambda$ first becomes dynamically important.  This redshift distribution
allows an opportunity to differentiate between gas evolution and
cosmological evolution of the abundance. This is shown in 
the left panel of Figure \ref{dndz_cd},
showing the redshift distributions of all clusters with a
central decrement above a given value, for different gas evolution
histories and cosmologies. Shown in the right panel of 
Figure \ref{dndz_cd} is the distribution
of all clusters with a total flux density above a given threshold
for two different cosmologies. The total flux density is 
almost independent of the gas evolution history, and is thus a very
clean probe of cosmology.

Two things to note in Figure 3 are 1) the large number of 
expected clusters, and 2) the sensitivity of a small beam
experiment to the gas evolution of clusters. The first point
is interesting since the cluster mass threshold used for
the plots have been selected to be within reach of 
planned SZ survey experiments. The importance of the second
point will be shown in the next section where its effect on
the angular power spectrum is discussed.

\begin{figure}
\epsfysize = 4.5 in
\centerline{\epsfbox{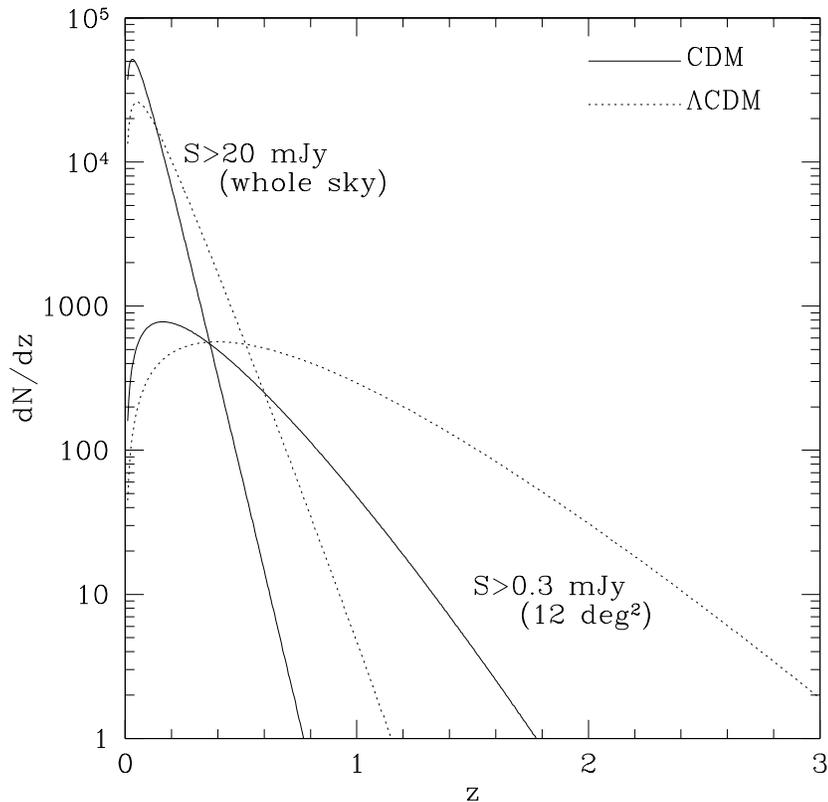}}
\caption{Expected number counts for two cosmologies for a low-sensitivity
full-sky SZ survey such as PLANCK compared with possible upcoming 
ground-based surveys as a function of redshift.}
\label{dn_surveys}
\end{figure}

CMB experiments are expected to, in some ways, act as SZ cluster
surveys. It is expected that satellite experiments should yield
a large catalog of clusters. This will provide excellent information
on the local abundance of clusters, but the catalog will not have
much information on the redshift evolution. To underscore this
point, we have compared expected cluster catalogs for full sky
coverage with a high flux cut-off with a 12 deg$^2$ survey with much better
sensitivity in Figure \ref{dn_surveys}.
This plot was done by assuming self-similar evolution, but
it can be seen in Figure \ref{dndz_cd}  that gas evolution does not have
much of an effect on this sort of plot.
The full-sky survey has a flux density cut-off of
20 mJy at 30 GHz, roughly comparable to the best expected performance
of MAP or PLANCK. The high-sensitivity survey is given a flux density
cut-off of 0.3 mJy, which may be possible with upcoming planned 
ground-based SZ surveys. 
While the large survey will yield more clusters, there is
very little information beyond the local cluster abundance, which is
already constrained by X-ray surveys (e.g., Viana and Liddle 1999). It
is the evolution of the cluster abundance which is most important for
cosmology. 

\subsection{Angular Power Spectrum}
\label{sec:angular}

The power spectra of the thermal and kinetic effects are practically
identical, with the main difference being an offset in amplitude by a factor
of $\sim 10-20$. The power spectrum is easy to understand. At large angular
scales (low $\ell$), clusters are unresolved, acting like point sources. The
angular Fourier transform of a point source has a flat amplitude, thus
the $c_\ell$ spectrum is flat. 
At small angular scales (large $\ell$), the clusters are resolved
and the structure of the cluster is the important. If we take clusters to have
$\rho \propto 1/(r_c^2 + r^2)$, then the angular SZ profile has $\Delta T
\propto 1/\sqrt{\theta_c^2+\theta^2}$. The angular
Fourier transform of this is proportional to $\exp(-\theta_c \ell) /\ell$, 
giving $\ell^2 c_\ell
\propto \exp(-2 \theta_c \ell)$.
Thus, a plot of $\ell(\ell+1)c_\ell$ for a single cluster
would rise as $\ell^2$ at large angular scales, while
it is suppressed as an exponential at small scales. The transition between
these two regimes happens around the characteristic scale of a cluster, 
the core radius, leading to a peak in the power spectrum near $\ell \sim 2000$.

Both the amplitude and position of the peak 
of the angular power spectrum are sensitive to the chosen
cosmology and the evolution of cluster gas. The amplitude 
at low $\ell$ is set by the total abundance of SZ effect-inducing
clusters. The local abundance is fairly well constrained from X-ray
observations, but the evolution with redshift is sensitive to cosmology.
Since the SZ effect is
redshift-independent, high-redshift clusters will contribute to CMB
anisotropy as far back as they exist. This is particularly noticeable in 
the slope of the power spectrum at low $\ell$'s. A model with a cosmological
constant has a significant contribution from compact high-redshift sources,
which have angular Fourier transforms which remain constant out to higher
values of $\ell$. Thus, the $\Lambda CDM$ model stays close to a line going
as $\ell^2$ longer than the flat matter-dominated model.

\begin{figure}
\epsfysize = 4.5 in
\centerline{\epsfbox{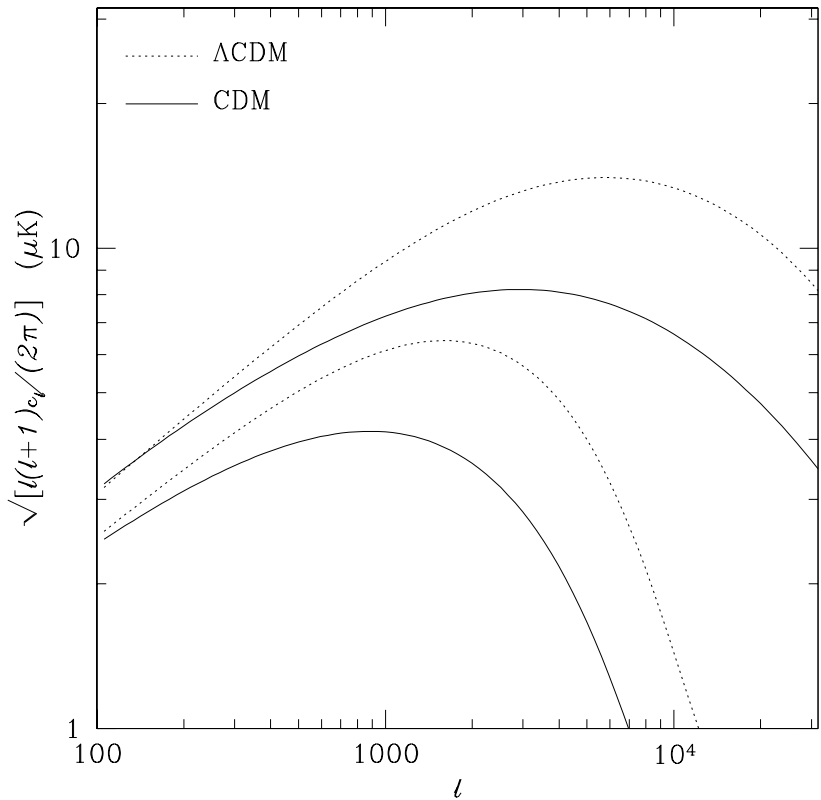}}
\caption{Angular power spectra $[{ \ell(\ell+1)c_\ell \over 2\pi}]^{1 \over 2}$ for two
cosmologies and the two gas evolution histories. The top curve 
for each cosmology corresponds
to self-similar evolution while the lower curve is for evolution with
a constant entropy core}
\label{c_l}
\end{figure}

The ratio of high-$\ell$ to low-$\ell$ contributions is effectively a measure
of a weighted average of angular core radii. Self-similar evolution 
leads to much smaller cores for higher
redshift clusters, and thus these power spectra have significantly
more signal at high $\ell$ than their isentropic counterparts. 
Low-density universes
have many more distant clusters (and thus more small-cored clusters) than
high density universes, giving more power at high $\ell$.
These two mechanisms, gas evolution and cosmological evolution of
the cluster abundance, can both change the average angular core radius  
and  lead to comparable effects in the SZ power spectrum, 
making a clean interpretation of an SZ power spectrum
difficult.

\begin{figure}
\epsfysize = 4.5 in
\centerline{\epsfbox{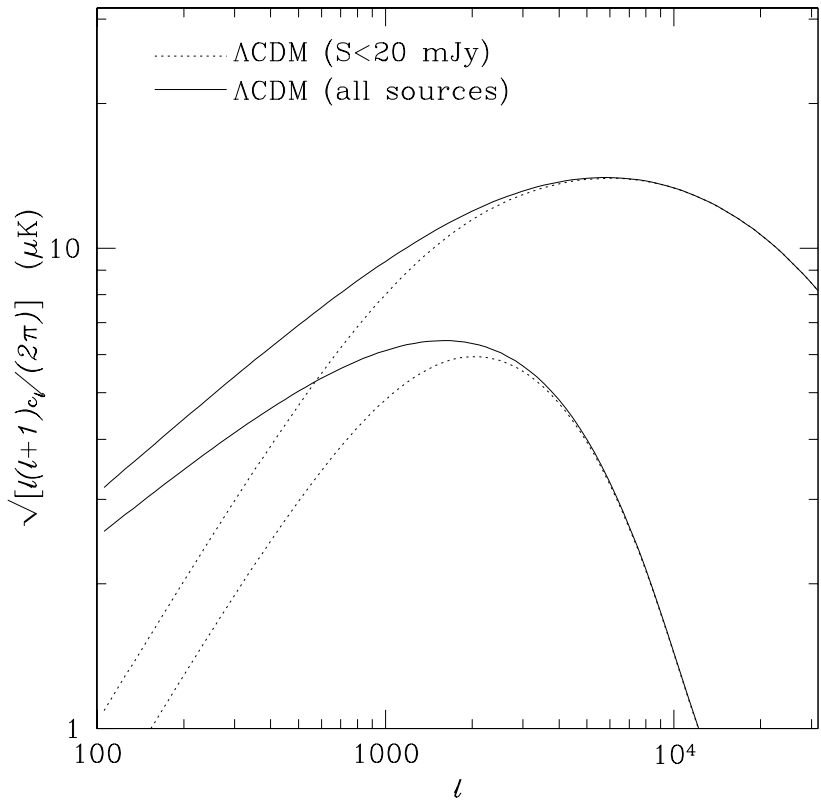}}
\caption{Angular power spectra $[{ \ell(\ell+1)c_\ell \over 2\pi}]^{1 \over 2}$ for a
$\Lambda CDM$ cosmology and the two gas evolution histories, 
showing the effect of removing all sources with a flux density 
greater than 20 mJy (at 30 GHz). The top curve (at high $\ell$) 
for each cosmology corresponds
to self-similar evolution while the lower curve is for evolution with
a constant entropy core}
\label{c_l_src}
\end{figure}

Shown in Figure \ref{c_l_src} is the effect of removing all bright
SZ sources above 20 mJy (at 30 GHz). While this is very effective
at low $\ell$, beating down the SZ signal by nearly a factor of 4, at
high $\ell$ this has little effect.
A large fraction of the signal at low $\ell$'s comes from these clusters.
The bulk of the high $\ell$ signal still comes from
distant clusters, undetected individually. This is problematic, since
it is exactly in this region where the SZ power spectrum starts to
dominate over the signal from the primary anisotropy. 

The thermal SZ effect does not approach the magnitude of the expected
CMB anisotropy until $\ell \ga 2000$. MAP has very little sensitivity at
this scale, and most current CMB experiments are not sensitive at this
scale either. However, many interferometric experiments have sensitivity
at this scale, and upcoming experiments such as PLANCK should definitely
pick up this SZ signal.

Added to this is uncertainty in the baryon fraction. 
Converting to BBN values would depress the expected
power from SZ clusters in a flat matter-dominated universe by
a factor of $0.2/\Omega_b$. While the evidence for BBN is strong,
the measured baryon fractions in clusters seem difficult to 
dismiss, making the choice of baryon fraction an important,
but difficult one in the case of $\Omega_M=1$.

\section{Summary}
\label{sec:summary}

The contribution to the CMB angular power spectrum by the SZ effect
from galaxy clusters is discussed. At low $\ell$ the primary CMB
anisotropy dominates the power spectrum. The SZ contribution rises
roughly as $\ell^2$ until $\ell \sim 2000$, where individual clusters
are resolved. At these $\ell$'s, the SZ contribution dominates that of
the primary CMB anisotropy.  Since the spectral distortion of the CMB
induced by the SZ effect is distinctly non-thermal, it can be
separated from intrinsic anisotropy for CMB observations which have
broad frequency coverage.

A measurement of the contribution to the CMB anisotropy power spectrum
from the SZ effect would provide a measure of the underlying
cosmology, independent of the values derived from the primary
anisotropy.  Combining large-scale non-targeted SZ effect surveys
currently under consideration and the SZ signal detected by CMB
experiments should give an independent determination of
$\Omega_m,\Omega_{\Lambda},{\rm and}\, \sigma_8$.

In order to complement CMB experiments, it would be useful to have SZ
effect surveys which can at least partially resolve clusters. 
As seen in Figures 3-6, it is important to have this 
structural information in order to constrain the evolution of the
cluster gas. Since typical core radii will be $1'$ or less, it would
be preferable to have a survey with a beam of this size or smaller.

The SZ effect should be more constructively considered as a low-redshift
source of secondary fluctuations in the CMB rather than simply as a foreground,
since the signature can be used as a direct constraint on cosmology.

\end{document}